\documentclass[final]
  {aipproc}

\layoutstyle{8x11double}

\begin{document}

\title{Ultrahigh Energy Cosmic Rays Detection}


\author{Carla Aramo}
{address={Istituto Nazionale di Fisica Nucleare - Sezione di Napoli\\
Complesso Universitario di Monte Sant'Angelo, Via Cintia, 80126 - Napoli}
}

\begin{abstract}
The paper describes methods used for the detection of cosmic rays with 
energies above 10$^{18}$ eV (UHECR, UltraHigh Energy Cosmic Rays).  
It had been anticipated there would be a cutoff in the energy spectrum 
of primary cosmic rays around 3$\cdot 10^{19}$ eV induced by their interaction 
with the 2.7 $^\circ$K primordial photons.
This has become known as the GZK cutoff.
However, several showers have been detected with estimated primary energy 
exceeding this limit.
\end{abstract}

\maketitle


\section{Introduction}
UHECRs form the tail of the cosmic-ray
spectrum, which extends from 1 GeV to beyond $10^{20}$ eV \cite{Watson}. 
Their energy is equivalent
to that of a tennis ball moving at 100 km/h and their flux is about once a 
year every 100 km$^2$ of the earth's surafce.
Because of their rarity we know relatively little
about them; in particular, we do not understand how or
where these particles gain their remarkable energies.
The prediction of the existence of the 
Greisen-Zatsepin-Kuzmin (GZK) cut-off \cite{GZK}
for particles with energy above 
3$\cdot 10^{20}$ eV has been faulted with the detection of several 
showers generated by a primary particles with energy well beyond 
$10^{20}$ eV \cite{Cris}.

In this paper methods used for UHECR detection will be described.
Moreover first results of P. Auger Observatory (PAO) will be shown.
This experiment \cite{Auger} has been conceived to measure
the properties of the highest-energy cosmic rays with
unprecedented statistical precision. 
The PAO detects ultra-high energy cosmic rays by 
implementing two complementary airshower techniques:
the combination of a large ground array and fluorescence detectors.
This hybrid observation of events allows a rich variety of measurements 
on a individual shower, providing much more
information than with either detector alone.\\
The complete observatory
will consist of two instruments, constructed in
the northern and southern hemispheres, each covering
an area of 3000 km$^2$.

\section{Detection techniques}

Currently UHECRs are investigated using two different detection methods. 
The first method consists on the distribution of a number of particle counters
across a large area allowing detection of particles
which survive to the detection level.
The other method exploits the excitation of nitrogen
molecules by the particles in the shower and the associated
fluorescence emission of light in the 300-400 nm
band. The light is detected by photomultipliers and the
profile of the shower in the atmosphere can be inferred
rather directly.

\subsection{UHECR detection with ground arrays}

An air shower produces a large number of particles
spread out over a large area at the observation
level. Particles are detected with an array of detectors
deployed over an appropriate area of many square
kilometers. The separation of detectors is typically many
hundreds of meters. The density of charged particle
and their arrival time are measured at each detector location.
These informations allow the reconstruction of shower axis and 
shower core (the impact point of the axis on the ground) by fitting the 
station signal size to expected lateral distribution function (LDF). 
The primary energy is estimated by a local charged particle density at 
fixed distance from the core in meters $S$(core distance) 
which depend by the array size. For example in the AGASA array was used 
$S(600)$ \cite{s600,Agasa} and in the PAO is used $S(1000)$ \cite{AugerIcrc}.
All of the arrays built to detect cosmic rays
above 10$^{19}$ eV have been located between 800 g cm$^{-2}$
and sea level. This is appropriate, as the average maximum
depth of showers of these primaries is about 750
g cm$^{-2}$ and it is effective to study showers close to or
beyond shower maximum.
The shower disc has a thickness that increases
from a few nanoseconds close to the shower core up to
several microseconds at distances beyond 1 km. 
The accuracy of the timing measurement is only one of the factors that
limit the directional precision: a second is the area of the
detector.
With giant arrays the arrival direction has been measured to
an accuracy of between 0.5$^\circ$ and 5$^\circ$.

\subsection{Detection with fluorescence detectors}

\begin{figure}
  \includegraphics[height=.265\textheight]{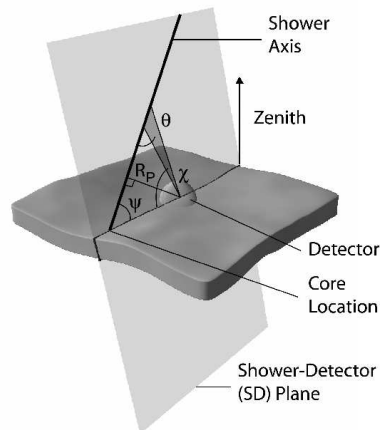}
  \caption{The geometry of the reconstruction for a monocular air 
fluorescence detector}
\label{SDP}
\end{figure}

The first successful implementation of the fluorescence
technique was obtained by the Fly's Eye group \cite{Bird}. 
The fluorescence detector follows the trajectory of
an extensive air shower and measures the energy dissipated
by shower particles in the atmosphere that acts as an air
calorimeter of more than 10$^{10}$ tons. For this purpose, the
whole sky is viewed by many segmented mirrors focusing the collected 
fluorescence light emitted isotropically
along the trajectory of the shower on a photomultiplier matrix.
Correlation between the light intensity and light arrival time detected in
each PMT provides unambiguous information on energy released and shower path 
in the atmosphere.  
The shower detector plane (SDP), defined in Figure \ref{SDP}, is
constructed from sequence of hit photomultipliers.
Then the distance to the shower axis (impact parameter)
Rp and incident angle $\psi$ in the plane are determined by
fitting the time sequence of several photomultiplier signals. 
Once the track geometry is determined, the number
of photons N$_\gamma$ received by a photomultiplier is calculated.
Since there is also contamination from direct and scattered
Cerenkov light, the longitudinal size N$_e$(x) of the
extensive air shower for each angular bin is calculated
via an iterative process to remove those contributions
which depend upon the viewing angle between the
pointing direction of the photomultiplier and the shower
axis. The resultant photoelectrons are directly proportional
to the number of charged particles in the angular
bin. From the integration, $\int$ N$_e$(x)dx , the total track
length is estimated.
If a shower is seen simultaneously by two fluorescence
detectors (stereo event), a shower detector plane for each one can be
determined and the intersection of these planes defines
the shower trajectory without timing information. The
total track length can also be determined independently
by each detector.

\subsection{AGASA and HiRes Experiment}

\begin{figure}
  \includegraphics[height=.26\textheight]{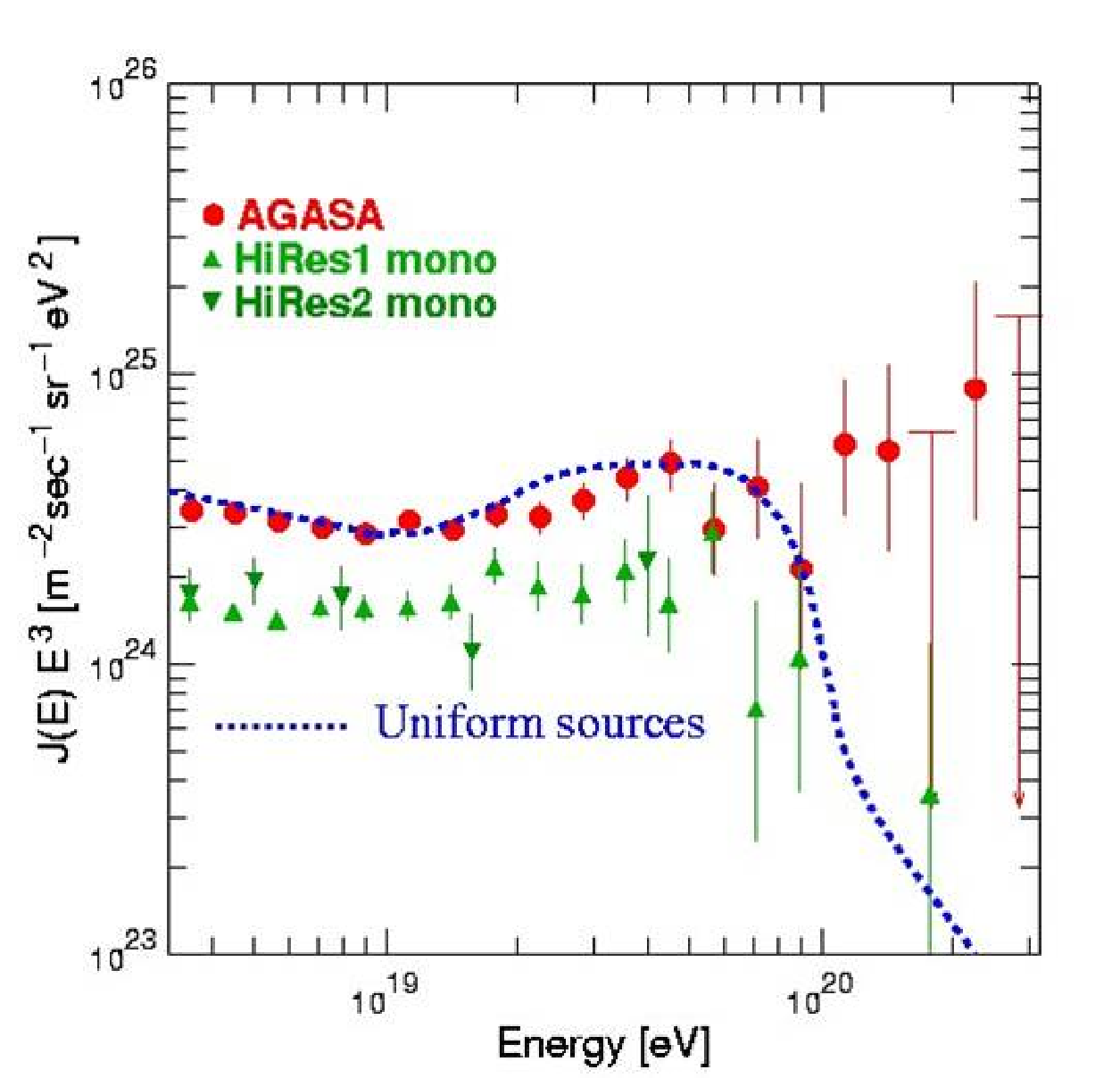}
  \caption{UHECR energy spectrum}
\label{agasa}
\end{figure}

As an example of the two techniques above described AGASA \cite{Agasa2} and 
HiReS \cite{HiRes} Experiment are discussed.
AGASA was operated at the Akeno Observatory (Tokyo), the operation had been 
started in February 1990 and was closed in January 2004 with an $\sim$ 95\%
overall live ratio. 
In an $\sim 100$ km$^2$ area 111 detector stations were deployed. 
Each station was equipped with 2.2 m$^2$ surface detector
with a 5 cm-thick  scintillator viewed by a PMT. 
At 27 southern stations muon detectors were built. They consisted 
of 14--20 proportional counters aligned below an absorber 
(30 cm-thick iron or 1 m-thick concrete: 0.5 GeV threshold energy for 
vertical incidence).
The primary energy was estimated by a local charged particle density at 
600 m from the core, known as $S(600)$:
$E [{\rm eV}] = 2.03 \times 10^{17} \cdot S(600)$.
The error of primary energy determination is 
$\pm 30\%$ at $10^{19.5}$eV and  $\pm 25\%$ at
$10^{20}$eV.
The exposure of AGASA is almost constant above $\sim 7\times 10^{18}$ eV  and
is $5.8\times 10^{16}$ m$^2$ sr s. 
Figure \ref{agasa} shows the energy spectrum of UHECRs above 
$10^{18.5}$~eV \cite{err}.  The vertical axis denotes 
the differential flux multiplied by ${E}^3$.  Poisson bounds are 
given at a confidence level (CL) of 68\%. Upper limits are given 
at a 90\% CL. The dashed curve represents the expected flux by 
the GZK hypothesis for the uniform source distribution \cite{yat}. 
The most noticeable feature is the observation of cosmic rays beyond  the GZK
cutoff energy. Eleven events were detected above
$10^{20}$ eV against the expected $\sim 1.9$ events \cite{Agasa2}.\\
HiRes Experiment have two detectors located atop desert mountains 
in west central Utah.  
The detectors consist of mirrors that collect fluorescence
light and focus it on arrays of 256 hexagonal photomultiplier tubes
(PMT's). Each PMT subtends about one degree of sky.  The HiRes-I
detector consists of 21 mirrors arranged to look from 3 to 17 degrees
in elevation and almost 360 degrees in azimuth.  The HiRes-II
detector, located 12.6 km SW of HiRes-I, consists of 42 mirrors, which
cover 3 to 31 degrees in elevation and almost 360 degrees in azimuth.
The spectrum of the HiRes-I and HiRes-II detectors, observing in
monocular mode are shown in Figure \ref{agasa} where a marked deficit 
of events above $10^{19.8}$ eV can be noted.  
This energy is the threshold for pion
production in interactions between cosmic ray protons and the average
photon of the CMBR; i.e., the deficit occurs at the energy of the GZK
cutoff. On the contrary the results of the AGASA experiment seem to indicate 
that the spectrum continues above the ankle at a constant power law.
To test whether HiRes data are consistent with this interpretation of
the Agasa result, the HiRes data were fit, from the ankle to the pion
production threshold, to a power law, then continue the power law to
higher energies.  This tests the hypothesis that the GZK cutoff is
absent, as the AGASA data seem to show.  The power law index is $2.8
\pm 0.1$.  If the cutoff were absent HiRes expect to see 29.0 events 
above $10^{19.8}$ eV, while only 11 event have been seen.  
The Poisson probability of seeing 11 or fewer events,
with a mean of 29.0, is $1 \times 10^{-4}$ \cite{HiRes}.  

\section{Pierre Auger Observatory}

The PAO was designed to observe, in coincidence, 
the shower particles at ground and
the associated fluorescence light generated in the atmosphere. 
This is achieved with a large array of water
Cherenkov detectors coupled with air-fluorescence detectors (eyes) 
that overlook the surface array. 
It is not simply a dual experiment. Apart from important cross-checks 
and measurement redundancy, the two techniques
see air showers in complementary ways. The surface detector (SD) measures the 
lateral structure of the shower
at ground level, with some ability to separate the electromagnetic 
and muon components. On the other hand,
the fluorescence detector (FD) records the longitudinal profile of the 
shower during its development through the atmosphere.
A hybrid event is an air shower that is simultaneously detected by the 
fluorescence detector and the ground
array. Data are recovered from both detectors whenever either system is 
triggered.
The Observatory started operation in hybrid production mode in January, 2004. 
Surface stations have a 100\% duty cycle while fluorescence eyes can only 
operate on clear moonless nights. Both surface and fluorescence
detectors have been running simultaneously 14\% of the time. 
The SD accumulated aceptance is larger than 1600 km$^2$ sr yr while the FD
aceptance is $\sim$ 14\% of the SD due to the limited duty 
cycle \cite{Matthews}.  
The number of hybrid events represents 10\% the statistics of the surface 
array data. 
The southern site of the Pierre Auger Cosmic Ray Observatory in Argentina 
now covers an area of approximately 1500 km$^2$ with an explosure of 1750 
km$^2$ sr yr and a full efficiency above 3 EeV for zenith angles less 
than 60$^\circ$ \cite{Parizot}.
Two of the Auger fluorescence detector sites (Los Leones and Coihueco) 
have been operating in a stable manner since January 2004 and a third site 
(Los Morados) began operation in March 2005.

\subsection{The Hybrid Performance and Measurements}

\begin{figure}
  \includegraphics[height=.37\textheight]{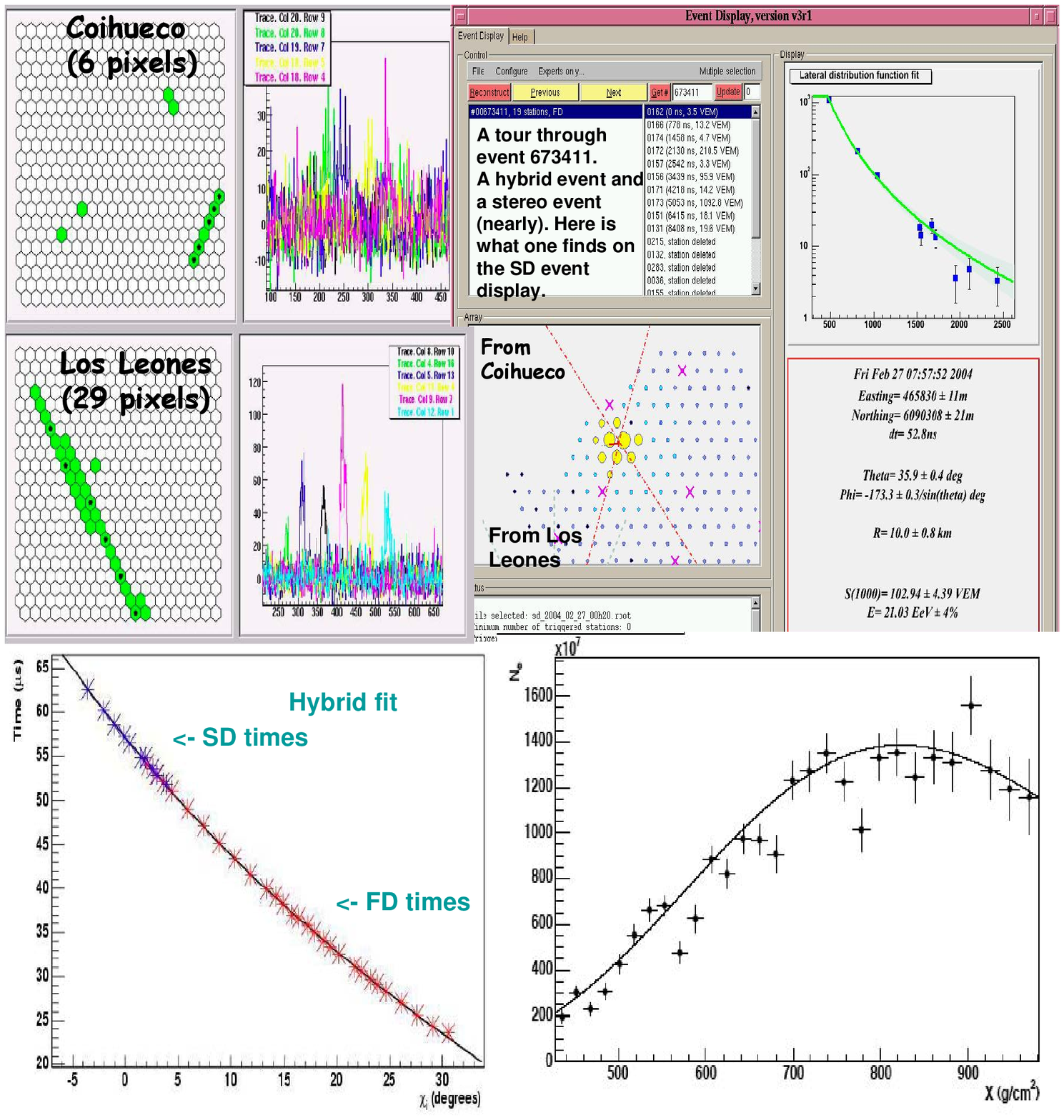}
  \caption{Example of hybrid stereo event (21 EeV). {\bf Top}: 
at left, green line 
is the hit PMTs in FD camera of Coihueco eye (up) and Los Leone eye (down)
with on the right the PMT signals of marked pixels.
At right the SD informations: on 
the left the core position (10 km) as cross point of two lines that are the SDP
from Coihueco and Los Leones and on the right the LDF function. {\bf Down}: 
at left the hybrid time fit is shown with blue (red) points related to 
tanks (pixels in the FD camera). On the right the
number of electrons as function of slant depth in the atmosphere is shown. The
solid line is the Gaisser-Hillas function that best fit the data.}
\label{hybrid}
\end{figure}

A hybrid detector has excellent capability for studying the highest energy 
cosmic ray air showers. Much of its capability stems from the accurate 
geometrical reconstructions it achieves. Timing information from even
one surface station can much improve the geometrical reconstruction of a
shower over that achieved using only eye pixel information. 
The reconstruction accuracy is better than the ground
array counters or the single eye could achieve independently. 
A core location resolution of 50 m is achieved. The resolution
for the arrival direction of cosmic rays is 0.6$^\circ$ \cite{Bonifazi}.
Due to the much improved angular accuracy, the hybrid data sample is ideal 
for anisotropy studies and, in particular, for point source searches. 
The combination of the air fluorescence measurements and particle detections 
on the ground provides an energy measurement almost independent of air shower 
simulations. The fluorescence measurements determine
the longitudinal development of the shower, whose integral is proportional 
to the total energy of the electromagnetic particle cascade. 
At the same time, the particle density at any given distance from the core 
can be evaluated with the ground array.
It is important to note that both techniques have
different systematics, and results are preliminary at this stage while 
the Observatory is under construction.
The hybrid analysis benefits from the calorimetry of the
fluorescence technique and the uniformity of the surface detector aperture.
Operation started in January, 2004 and over 16000 hybrid events have been 
successfully reconstructed up to now \cite{Mostafa}. 
An example of hybrid and stereo (two FD eyes) event with an energy of 
$\sim$ 21 Eev is reported in Figure \ref{hybrid}. 

\subsection{The first AUGER energy spectrum and Anisotropy Studies}

\begin{figure}
  \includegraphics[height=.27\textheight]{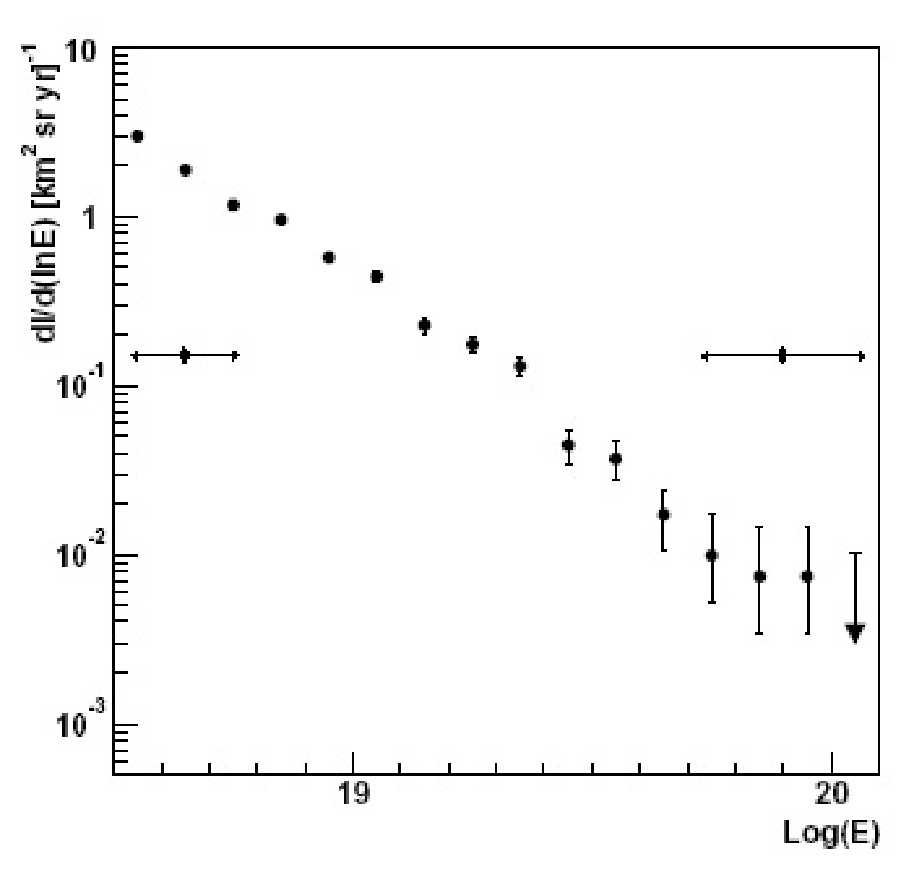}
  \caption{Estimated Auger energy spectrum \cite{Sommers}. Error bars
on points indicate statistical uncertainty. 
Systematic uncertainty is indicated by double arrows
at two different energies.}
\label{spectrum_auger}
\end{figure}

The methods to calculate the cosmic ray energy 
spectrum are simple and robust, exploiting the combination of 
FD and SD. The methods do not rely on detailed numerical simulation 
or any assumption about the chemical composition.
The spectrum in Figure \ref{spectrum_auger} is only a first estimate and 
is related with data from 1 Jan 2004 through 5 Jun 2005 \cite{Sommers}. 
Events are included for zenith angle 0-60$^\circ$ and for energies above 
3 EeV, in total 3525 events.
It has significant systematic and statistical uncertainties. The indicated 
statistical error for each point comes directly from the Poisson uncertainty 
in the number of measured showers in that logarithmic energy bin. 
There is larger systematic uncertainty in the conversion of S(1000) to energy. 
Part of that comes from the FD energies themselves. 
The accuracy is limited by the available statistics, and the total systematic 
energy uncertainty grows with energy: from 30\% at 3 EeV to 50\% at 100 EeV. 
This uncertainty is indicated 
by horizontal double arrows in Figure \ref{spectrum_auger}, and a 
10\% systematic uncertainty in the exposure is indicated by vertical arrows. 

The Auger data have been analyzed to search for excesses of events 
near the direction of the galactic center in several energy ranges around 
EeV energies \cite{Selvon}. In this region the statistics accumulated by the 
Observatory are already larger than that of any previous experiment. 
Using both the data sets from the surface 
detector and hybrid data sets any significant excess is find. 
These results do not support the excesses reported by AGASA 
and SUGAR experiments. An upper bound on the flux of cosmic rays arriving 
within a few degrees from the galactic center in the energy range 
from 0.8-3.2 EeV is set: $\Phi_s<2.8\,\, \xi\, 10^{-15}\, \rm m^{-2}\,s^{-1}$ 
@ 95\% CL with $\xi$ in [1.0-4.0] \cite{Selvon}. 
Also the search for correlations of cosmic ray arrival directions with the 
galactic plane and with the super-galactic plane at energies in the range 
1-5 EeV and 12 above 5 EeV found no significant excess \cite{Selvon}.
\vspace{-0.3cm}
\begin{theacknowledgments}
I wish to thank all the P. Auger Collaboration, in particular the Napoli group.
I thank the organizers of IFAE 2005 for an exciting workshop and for 
their hospitality. 
\end{theacknowledgments}

\end{document}